\documentclass[fleqn,usenatbib]{mnras}

\usepackage{txfonts}
\usepackage[T1]{fontenc}

\DeclareRobustCommand{\VAN}[3]{#2}
\let\VANthebibliography\thebibliography
\def\thebibliography{\DeclareRobustCommand{\VAN}[3]{##3}\VANthebibliography}

\usepackage{graphicx}	
\usepackage{amssymb}	

\title[GRAVITY data of the post-Red Supergiant IRC+10420]{Tracing a decade of activity towards a yellow hypergiant. The spectral and spatial morphology of IRC+10420 at au scales.}

\author[E. Koumpia et al.]{
Evgenia Koumpia,$^{1,2}$\thanks{E-mail: ekoumpia@eso.org}
R. D. Oudmaijer,$^{2}$
W.-J. de Wit,$^{1}$
A. Mérand,$^{3}$
J. H. Black,$^{4}$ 
\newauthor and K. M. Ababakr$^{5}$
\\
$^{1}$ESO Vitacura, Alonso de Córdova 3107 Vitacura, Casilla, 19001, Santiago, Chile \\
$^{2}$School of Physics \& Astronomy, University of Leeds, Woodhouse Lane, LS2 9JT, Leeds, UK\\
$^{3}$ESO, Karl-Schwarzschild-Straße 2, 85748 Garching bei München, Germany\\
$^{4}$Department of Space, Earth, and Environment, Chalmers University of Technology, Onsala Space Observatory, 43992 Onsala, Sweden \\
$^{5}$Erbil Polytechnic University, Kirkuk Road, Erbil, Iraq}

\date{Accepted XXX. Received YYY; in original form ZZZ}

\pubyear{2022}

\begin{document}
\label{firstpage}
\pagerange{\pageref{firstpage}--\pageref{lastpage}}
\maketitle

\begin{abstract}
The fate of a massive star during the latest stages of its evolution is highly dependent on its mass-loss history and geometry, with the yellow hypergiants being  key objects to study those phases of evolution. We present near-IR interferometric observations of the famous yellow hypergiant IRC +10420 and blue spectra taken between 1994-2019. Our 2.2~$\mu$m GRAVITY/VLTI observations attain a spatial resolution of $\sim$5 stellar radii and probe the hot emission in the K-band tracing the gas via Na {\sc i} double emission and the Br$\gamma$ emission. The observed configurations spatially resolve the 2.2~$\mu$m continuum as well as the Br$\gamma$ and the Na {\sc i} emission lines. Our geometric modelling demonstrates the presence of a compact neutral zone (Na {\sc i}) which is slightly larger than the continuum but within an extended Br$\gamma$ emitting region. Our geometric models of the Br$\gamma$ emission confirm an hour-glass geometry of the wind. To explain this peculiar geometry we investigate the presence of a companion at 7-800 au separations and find no signature at the contrast limit of our observations (3.7 mag at 3$\sigma$). We also report an evolution of the ejecta over a time span of 7 years, which allows us to constrain the opening angle of the hour-glass geometry to be $<$10$^\circ$. Lastly, we present the first blue optical spectra of IRC +10420 since 1994. The multi-epoch data indicate that the spectral type, and thus temperature, of the object has essentially remained constant during the intervening years. This confirms earlier conclusions based on mostly photometric data that following  an increase in temperature of 2000 K in less than two decades prior to 1994, the temperature increase has halted.  This suggests that this yellow hypergiant has "hit" the White Wall in the HR-diagram preventing it from evolving blue-wards, and will likely undergo a major mass-loss event in the near future.  
\end{abstract}


\begin{keywords}
stars: evolution -- techniques: interferometric -- stars: individual: IRC+10420 -- stars: mass-loss
\end{keywords}



\section{Introduction}

Massive stars ($>$8~M$_\odot$) are among the most influential objects in the universe, as they provide the primary energy budget in galaxies, they produce heavy elements enriching the chemical composition of the interstellar medium, and even trigger star formation in their immediate surroundings \citep{Lee2007}. Many important open issues related to the final stages of stellar evolution can only be addressed once we comprehend the properties of circumstellar material close to the star. For massive stars, which eventually explode as core-collapse supernovae (SN), understanding the mass-loss geometry, rate and history is particularly important \citep{Heger1998}. Mass-loss events impact both the angular momentum evolution and the final mass, and therefore, the fate of the individual massive star, while creating the circumstellar environment with which SN ejecta will interact \citep{Patat2011,Moriya2014}. However, the exact mechanism shaping the ejecta around evolved massive stars remains uncertain.

A suitable class of stars to study mass-loss events are post-red supergiants (post-RSG), but only a few such objects are known. The A- to K-type yellow hypergiants (YHGs), such as IRC+14020, IRAS 17163-3907, and HD179821, show evidence of circumstellar dust and high mass-loss rates \citep[][and references therein]{Oudmaijer2009,Humphreys2020}. Therefore, these objects are great post-RSG candidates making excellent laboratories to study the mass-loss events that take place during the post-RSG evolution. In those phases, the prominent mechanisms of mass-loss include pulsational-driven  and  line-driven  mass-loss \citep[e.g.,][]{Lobel1994,Vink2002}. In a line-driven  mass-loss  scenario,  the  interacting  winds  are driven by radiation pressure in spectral lines, and any change in radiative acceleration affects both the wind velocity and  mass-loss  rates  (e.g. the bistability  jump, see discussion in \citealt{Koumpia2020}).  Still  it  is  very  challenging  to  distinguish  between  those  mechanisms, and improved measurements of the mass-loss process as traced by the dust and gas emission are required.

 In this study we focus on IRC+10420, an object that is among the brightest IR sources in the sky. The central star is as massive as 25-40 M$_{\odot}$, and is most likely to be in the stage where it is evolving quickly towards the blue part of the Hertzsprung–Russell diagram \citep[see e.g.,][]{Meynet2015}, and its final fate as an SN. IRC+10420 also shows evidence of episodic mass loss events \citep[e.g.,][]{Oudmaijer1994,Gordon2019} and discrete ejection regions \citep{Tiffany2010}. In particular, it is found to have undergone variable mass-loss of two distinct mass-loss periods constituting of a high mass loss rate of 2$\times$10$^{-3}$ M$_\odot$/yr  until  about  2000  yr  ago,  followed  by  a  significant  decrease  of  about  an  order  of  magnitude  in  the  recent past \citep{Shenoy2016}. This result was based on a simultaneous fit of the spectral energy distribution (SED) with a spatially resolved image taken with MIRAC4/MMT at 11.9 $\mu$m. Optical high spectral resolution observations traced IRC+10420 for more than a decade (2001-2014) and revealed a rather stable kinematic behaviour of IRC+10420's atmosphere \citep{Klochkova2016}.  
 
 The object has also been seen to be evolving in the HR diagram from the red to the blue. Its spectral type was determined to be that of an F8-G0I supergiant in 1973 \citep{Humphreys1973}, while no hydrogen recombination line emission was observed at the time \citep{Thompson1977}.
 \citet{Oudmaijer1994} discovered hydrogen Br$\gamma$ line emission in the object and traced  the onset of the hydrogen
recombination line emission of the object back to a period between
1984-1986. These dates coincide with the sudden change in $V - K$
colour, which \citet{Oudmaijer1996} interpreted as  the result of
an increase in temperature of the object. The latter was validated by
the spectral classification of the star as being mid-A supergiant by
\citet{Oudmaijer1998} which was confirmed by \citet{Klochkova1997}
whose atmospheric modelling led to a temperature of 8500 K for the
star based on the same data. All the evidence seems to indicate that the object increased in temperature of $\sim$2000 K over two decades. The stable photometry since seems to indicate that the temperature evolution has halted \citep{Patel2008}, while the emission line spectra also seemed to imply that the
temperature evolution had ceased or at least slowed down
\citep{Klochkova2016}. Determining the spectral type of the object is not trivial; the star is bright in the red wavelength range, which is dominated by emission lines, rendering any spectra unusable for spectral typing purposes. IRC +10420 is much fainter in the blue, but it displays a rich  absorption spectrum suitable for determining its spectral type.  Thus far, the only determination of the object's spectral type and thus temperature after its evolution from being an F8-G0I supergiant
in 1973 was the 1994 blue optical spectrum.  In this paper we present multiple blue spectra spanning 25 years since 1994 and investigate whether the spectral type has changed significantly or not. 

To trace the mass-loss history via the ejected dusty environment surrounding evolved massive stars, it is crucial to observe them in the near-infrared (NIR), as a large fraction of the emission coming from ejected dusty shells is thermal. In this paper, we focus on the K-band emission using GRAVITY/VLTI and investigate the spectral and spatial variability of IRC+10420 in the near-infrared by combining our findings with those of literature. 

\section{Methods, Observations}

\subsection{GRAVITY observations and data reduction}
\label{sec:obs} 

\begin{figure}
\begin{center}  
\includegraphics[scale=0.45]{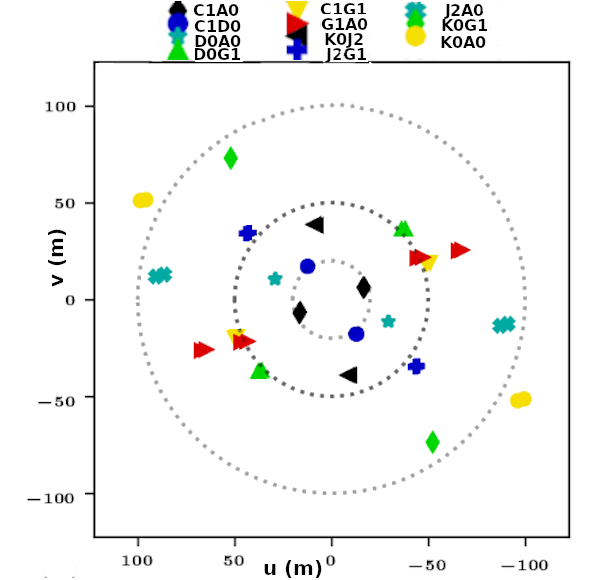} 
\end{center}
\caption{uv-plane coverage of VLTI/GRAVITY observations of IRC+10420.}
\label{fig:uvplane}
\end{figure}

IRC+10420 (RA = 19$^{h}$26$^{m}$48$^{s}$, Dec = +11$^{\circ}$21\arcmin16\arcsec.8 [J2000]) was observed during four nights in June 2016 with GRAVITY \citep{Gravity_Coll2017,Eisenhauer2011} on the VLTI using all four 1.8~m Auxiliary Telescopes (ATs). GRAVITY is an interferometer which operates in the K-band (1.99~$\mu$m-2.45~$\mu$m) at a range of spectral resolutions. The data were taken in high resolution (HR; R$\sim$4000) combined mode, which corresponds to a velocity resolution of $\sim$75~kms$^{-1}$. Two baseline configurations were used, the medium (A0-G1-D0-C1) and large (A0-G1-J2-K0), which provided a good uv-coverage (Figure~\ref{fig:uvplane}) and projected baseline lengths between $\sim$20~m and 130~m. At the observed configurations the highest angular resolution achieved is $\sim$1.7~mas at 2.2~$\mu$m. The physical scales to be resolved are down to $\sim$7~au ($\sim$5 R$_{*}$) at the distance of IRC+10420 (4.26~kpc; see Section~\ref{dist}). The shorter baselines covered a wide range of position angles (PAs) between 0 and 180$^{\circ}$ while the longer baselines ($>$ 100~m) covered position angles between $\sim$ 0 and 30$^{\circ}$. The observations were taken under good atmospheric conditions with the seeing being mostly below 1\arcsec and the coherence time at V between 3 and 7~ms. The technical overview is given in Table~\ref{gravity_tech}.   

\begin{table*}
\caption{Technical overview of the GRAVITY observations on ATs towards IRC+10420 at the start of each run. The individual exposure time is noted as DIT. The coherence time in the visible $\tau_{coh}$, the seeing, and the visibilities of the continuum are also reported.}
\small
\centering
\setlength\tabcolsep{2pt}
\begin{tabular}{c c c c c c c c c c c c c }
\hline\hline
Configuration & Date & Station & Baseline & PA & DIT & $\tau_{\rm coh}$ & Seeing & V$_{\rm cont}$  \\ & & & (m) & ($^\circ$) & (s) & (ms) & (arcsec) &  \\
\hline\hline
A0-G1-J2-K0 & 2016-06-18 & K0J2 & 38.6 & 94.0 & 10 & 4 & 0.73 & 0.849$\pm$0.006   \\
& & K0G1 & 82.6 &  55.3 &  &  &  &  0.658$\pm$0.009 \\
& & K0A0 & 127.8 & 18.1 &  &  &  & 0.509$\pm$0.008  \\
& & J2G1 & 57.8 & 30.5 &  &  &  & 0.75$\pm$0.02  \\
& & J2A0 & 124.2 & 0.5 &  &  &  & 0.54$\pm$0.01  \\
& & G1A0 & 79.6 & 159.2 &  &  &  & 0.88$\pm$0.01  \\
A0-G1-J2-K0 & 2016-06-20 & K0J2 & 39.7 & 78.0 & 10 & 8 & 0.41 & 0.77$\pm$0.01  \\
& & K0G1 & 89.8 &  54.5 &  &  &  & 0.50$\pm$0.01  \\
& & K0A0 & 111.4 & 27.3 &  &  &  &  0.39$\pm$0.02 \\
& & J2G1 & 55.7 & 37.9 &  &  &  & 0.67$\pm$0.02  \\
& & J2A0 & 91.6 & 7.7 &  &  &  & 0.68$\pm$0.03  \\
& & G1A0 & 51.6 & 154.9 &  &  &  &  0.767$\pm$0.001 \\
A0-G1-D0-C1 & 2016-06-21 & C1D0 &  21.4 & 54.1 & 10 & 4 & 0.69 &  0.721$\pm$0.003  \\
& & C1G1 & 54.0 &  158.9 &  &  &  & 0.815$\pm$0.002  \\
& & C1A0 & 18.0 & 158.9&  &  &  & 0.817$\pm$0.002  \\
& & D0G1 & 52.8 & 135.8 &  &  &  & 0.744$\pm$0.002  \\
& & D0A0 & 31.3 & 20.3&  &  &  & 0.710$\pm$0.002  \\
& & G1A0 & 72.1 & 158.9 &  &  &  & 0.805$\pm$0.002  \\
\hline\hline
\end{tabular}
\label{gravity_tech}
\end{table*}

HR~7648 (RA = 20$^{h}$00$^{m}$59$^{s}$, Dec = +08$^{\circ}$33\arcmin27\arcsec.8 [J2000]) served as a standard calibration object during all observations. The spectral type of HR~7648 is K5III, it is bright in the K band (2$^{m}$.2) and with a well-known uniform disc diameter of 1.96~mas \citep[JMMC SearchCal;][]{Bonneau2011}. 

The reduction and calibration of the observations was performed using the default parameters of the standard GRAVITY pipeline recipes as provided by ESO (version 1.0.5\footnote{The ESO pipeline for the GRAVITY data reduction can be downloaded at https://www.eso.org/sci/software/pipelines/gravity/}). The spectra were corrected for tellurics using the HITRAN models \citep{Rothman2009} via the PMOIRED \footnote{PMOIRED is a Python3 module which can be downloaded at https://github.com/amerand/PMOIRED} package, which uses synthetic atmospheric transmission models very similar to Molecfit \citep{Smette2015}. The use of synthetic transmission during telluric correction is found to deliver an improved solution compared to that of a standard star \citep{Ulmer2019}.

\subsection{K-band spectrum}

\begin{figure*}
\begin{center}  
\includegraphics[scale=0.25]{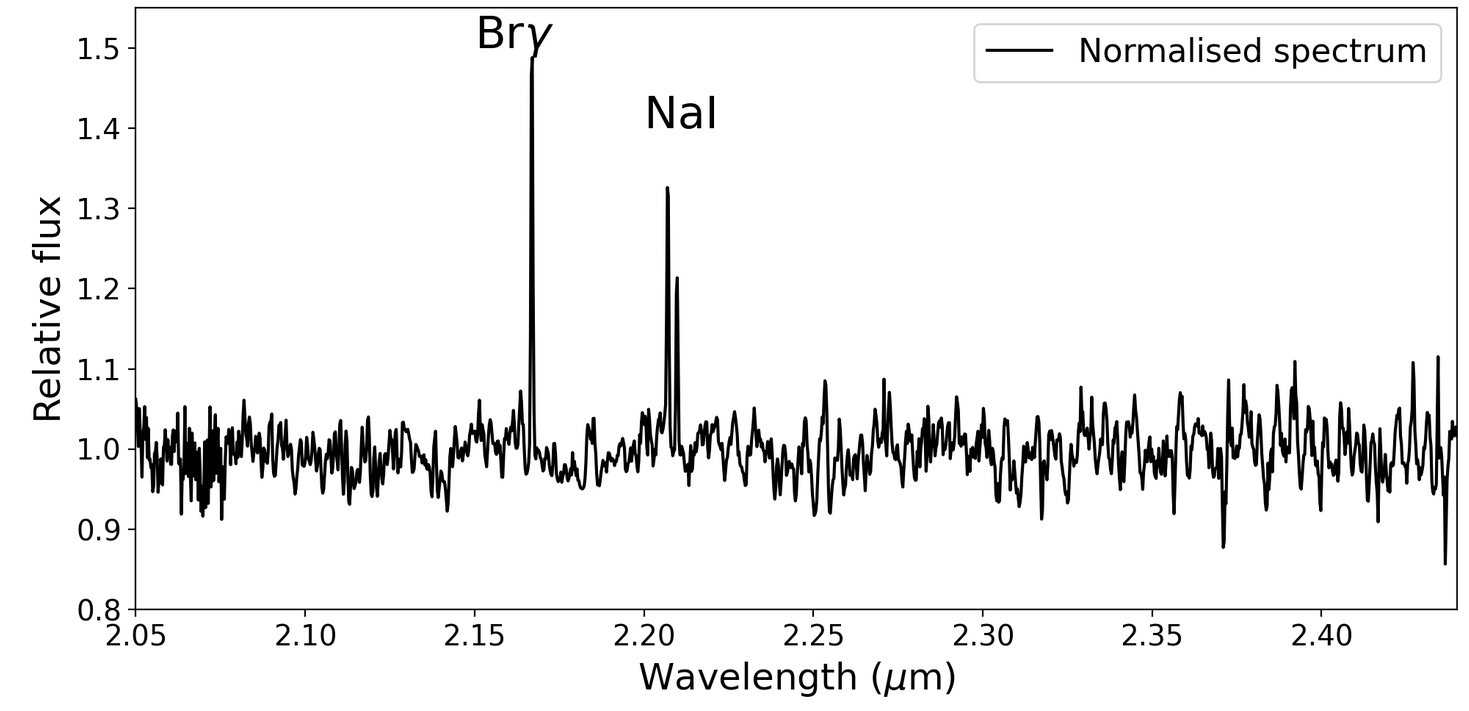} 
\end{center}
\caption{Normalised spectrum of IRC+10420 as observed with GRAVITY after telluric correction. The only prominent spectral features are the Br$\gamma$ and Na {\sc i} doublet emission.}
\label{fig:spectrum}
\end{figure*}

We present a new NIR spectrum of IRC+10420 as obtained with GRAVITY in 2016 (Figure~\ref{fig:spectrum}) after telluric correction and normalisation. The observed wavelength coverage (2-2.4~$\mu$m) contains the hydrogen recombination emission line at 2.167~$\mu$m (Br$\gamma$), and the Na {\sc i} 2.206~$\mu$m and 2.209~$\mu$m doublet also in emission. The observed spectral range covers the wavelength region where the CO bandhead is located but we do not observe the characteristic signals in absorption or emission, which were also not clearly detected in low resolution CRSP K-band spectra \citep[R$\sim$500;][]{Humphreys2002}. The K-band spectrum of IRC+10420 is very similar to the one previously observed towards another famous yellow hypergiant, IRAS 17163, with the exception of the Mg {\sc ii} 2.137~$\mu$m and 2.144~$\mu$m emission, which is only present towards IRAS 17163-3907, the central star of the "Fried Egg" nebula \citep{Koumpia2020}.

 We further investigate the spectral variability of IRC+10420 over a 7 year time interval, by comparing our more recent K-band spectrum around the Br$\gamma$ and Na {\sc i} emission lines with an X-Shooter spectrum of the object which was obtained in 2009 and presented in \citet{Oudmaijer2013} (hereafter OdW13). The line-to-continuum ratio of Br$\gamma$ is $\sim$1.5, which is comparable to the previous value of $\sim$1.6 reported using X-Shooter,  given the superior spectral resolution of the latter (R $\sim$ 10,000 versus 4,000). We also note that the two instruments are not characterised by the same spatial resolution or FoV (10 $\times$ larger for X-Shooter than GRAVITY). As a result, GRAVITY over-resolves the dust continuum, while X-Shooter traces a larger dust emitting area and captures the K-band excess emission. The strongest line of the Na {\sc i}  doublet, is the one at 2.206~$\mu$m and is characterised by a line-to-continuum ratio of 1.34.  
 
 The strongest line of the Na {\sc i} doublet is a bit broader ($\sim$100~kms$^{-1}$) compared to the FWHM of the Br$\gamma$ emission ($\sim$85~kms$^{-1}$). The blue component of the Na {\sc i} doublet is $\sim$ 40\% stronger than the red component. Our measurements are broader than those reported in OdW13 ($\sim$70~kms$^{-1}$) which can be explained given the GRAVITY spectral resolution ($\sim$ 75~kms$^{-1}$).
 
The EW of the Br$\gamma$ emission line in the GRAVITY spectrum (2016) is -3.4~\AA. Note that OdW13 reported an EW of -4.1~\AA~ (2009), \citet{Humphreys2002} -4~\AA~ (2000), while previous AMBER observations on the source show an EW of -6.7~\AA~ (2006, 2008). The EW measurements of the individual components of the Na {\sc i} doublet emission are -2.8~\AA~ and -1.5~\AA~, which are both smaller to what was previously reported (-3.7 and -2.9~\AA; OdW13). Our reported EWs are characterised by errors up to $\sim$ 10\%, while the reported differences among the studies are affected by instrumental contributions (i.e., the spatial and spectral resolution differ). Therefore, although the reported variations can be due to a real source variability, the above restrictions do not allow us to properly assess this hypothesis, for which higher spectral resolution and homogeneous spatial resolution observations are necessary.

\begin{table*}
\caption{Log of the spectroscopic observations}
\begin{small}
\begin{tabular}{llllc}
\hline
Date  & Instrument    & Wavelength range    & Exposure time & Spectral resolution \\
\hline
1994, 27, 28 July       & WHT/UES & 3800 - 10,000 \AA & 6.5h in blue & 10 kms$^{-1}$ \\
2006, 22 May, 16,17,18 June & ESO 2.2/FEROS & 3800 - 9000   \AA & 5.5h        & 5  kms$^{-1}$ \\
2009, 15 August  & VLT/X-Shooter  & 3000 - 24,000 \AA & 30m in blue & 30 kms$^{-1}$ \\
2013, 05 August  & ESO 2.2/FEROS  &   3800 - 9000   \AA & 6000s        & 5  kms$^{-1}$ \\
2019, 19 September & WHT/ISIS     &  3800 - 4600  \AA & 2700s        & 70  kms$^{-1}$ \\
\hline
\end{tabular}
\label{speclog}
\end{small}
\end{table*}

\subsection{Interferometric observables}

The interferometric observables as obtained with GRAVITY consist of the spectral information, calibrated visibilities, differential phases and closure phases in the entire range of 2.0-2.4~$\mu$m. In Figure~\ref{fig:vis_brg1_2_irc} we focus on the observables around the Br$\gamma$ and Na {\sc i} emission for the large configuration. The top panel shows the flux, followed by the visibility (second panel), differential phase (third panel) and closure phase (bottom panel) as a function of wavelength for all six baselines and four unit triangles. 

The visibilities around the Br$\gamma$ emission are characterised by a significant decrease in their value (up to 50\% at longest baselines) compared to the visibilities of the continuum (Table~\ref{gravity_tech}). The observed decrease in visibilities is about an order of magnitude larger than the corresponding errors (up to 5\% accounting for calibrator and transfer function uncertainties), and therefore, can be attributed to real geometrical effects. The observed trend is apparent at all available baseline lengths and is suggestive of a larger Br$\gamma$ emitting region compared to the continuum at all traced position angles (0-160$^{\circ}$). In addition, our observations around the Br$\gamma$ emission also have decreasing differential phases at all baseline lengths and position angles of the observed configuration, while the closure phases decrease for two out of the 4 observed triplets compared to the continuum. The detected changes in differential phases suggest that the Br$\gamma$ emitting region is characterized by a different photocentre compared to the continuum emitting region. The detected changes in closure phases on the other hand, suggest a non-axisymmetric brightness distribution. 

Looking at the interferometric observables around the Na {\sc i} emission we observe a weak drop of the visibilities compared to the continuum, with the exception of the shortest baseline of $\sim$40~m which shows no drop. These interferometric findings suggest that Na {\sc i} doublet emission originates from a larger region compared to the continuum, but smaller compared to Br$\gamma$ emission. In contrast to Br$\gamma$, the Na {\sc i} emission does not show significant changes in differential or closure phases compared to the continuum with the observed configuration,  which is suggestive of its symmetric nature with respect to the continuum, at least at the traced scales.

\subsection{Optical Spectroscopic Observations}

We have obtained blue optical spectra, covering the classical wavelength range
used for spectral classification, of IRC+10420 spanning a period
of 25 years. Although the object is bright at red
wavelengths (the $R$ band which covers H$\alpha$ is $\sim$ 9 mag.), it
is very reddened and blue spectra have not often been
published. Indeed, \citet{Oudmaijer1998}'s 1994 spectra which led to
the initial A-supergiant classification remain the only published such
data as of yet.

An overview of the data is provided in Table~\ref{speclog}.  The 1994
UES data taken at the 4.2m William Herschel Telescope, La Palma, and part of the 2009 X-Shooter data from ESO's 8m VLT were reported on in
\citet{Oudmaijer1998} and OdW13 respectively and more
details on the observations can be found there.  The 2006 and 2013
FEROS data obtained with the 2.2 m telescope were retrieved from the ESO archive and were already
pipeline reduced. FEROS data were taken on several nights in 2006, in order to increase the signal-to-noise these were combined after verifying that no obvious variations occured over the 1 month timespan. In much of the analysis hereafter, the blue spectra
have been rebinned to achieve a higher signal-to-noise ratio. Given
that the FEROS spectral resolution is far larger than that of
X-Shooter for example, the final difference in data quality is
modest. We note however, that for spectral typing purposes, the data
are of excellent quality. In 2019, the WHT/ISIS combination was used
with the R1200B and R1200R gratings on the blue and red arms
respectively. Here, we discuss the blue parts of the spectrum. 


\section{Results}

\begin{figure*}
\begin{center}  
\includegraphics[scale=0.34]{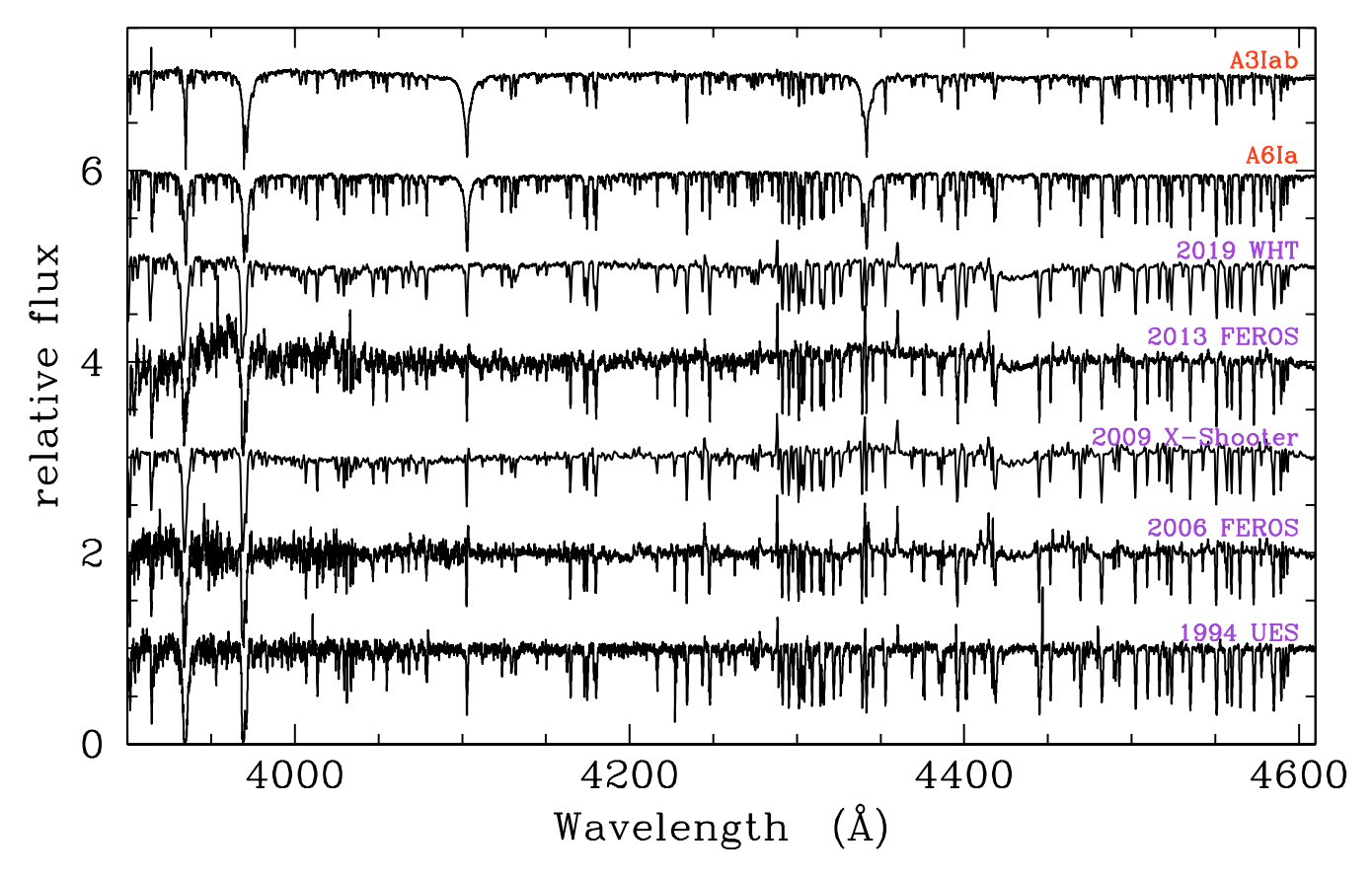}
\end{center}
\caption{Blue optical spectra of IRC+10420 obtained with different instruments/observatories. Data have been rebinned to improve the signal-to-noise, are continuum normalised and offset by increases of 1. For reference spectra of 2 spectral standard stars from \protect\citet{Bagnulo2003} are plotted.  The data demonstrate that this YHG has retained its spectral type over a time-span of 25 years.}
\label{fig:blue_spectrum}
\end{figure*}
\subsection{Spectral type}

%

Figure~\ref{fig:blue_spectrum} shows the 1994, 2006, 2009, 2013 and
2019  spectra in the wavelength range 3900 - 4600 \AA, which
is the classical wavelength range for spectral classification. For reference the spectra of two spectral standard stars bracketing the spectrum of IRC +10420 are shown at the top. These are taken from the high resolution catalog of \citet{Bagnulo2003}.   IRC +10420 is intermediate between both stars, in some instances (e.g. the lines around 4125~\AA) it seems more similar to the A3Iab supergiant while in other, more, instances (e.g. around 4175~\AA) it appears closer to the A6Ia object. These apparent discrepancies already illustrate the difficulty in assigning a spectral type to highly evolved supergiants  and signal the need for dedicated modelling to arrive at more quantitative values for the stellar parameters (cf. \citealt{Klochkova1997}).

Having said that, given that the data were taken with different instruments and
different telescopes at different spectral resolutions and
signal-to-noise ratios, it is remarkable how similar the data are at the various epochs. The variations do not appear to be large. When using the spectral type - temperature calibration of \citet{Gray2009}, we find that the A3 supergiant is $\sim$450~K hotter than the A6 supergiant. The spectral variations of IRC +10420 - even if present - indicate temperature changes that are significantly less than this.  We therefore conclude that there has  not, or hardly  been, a
temperature evolution of IRC +10420 between at least 1994, the
earliest spectrum and 2019, the most recent spectrum that we
obtained. This confirms earlier reports in the literature which are summarised in the Introduction, that the object had a
sudden increase in temperature of around 2000 K in the early 1980's to
then remain stable in temperature \citep{Oudmaijer1996,Patel2008}.



\subsection{Distance and luminosity}

\label{dist}

The accurate determination of the distance of IRC+10420 is crucial to confirm its class as a YHG, and to constrain its luminosity, stellar radius, as well as the physical sizes of both the 2.2~$\mu$m continuum and line emission. The distance inferred from various studies so far was based on various methods (radial velocity measurements, interstellar extinction, diffuse interstellar bands, Na~D absorption lines) and was found to be within the range of 3-5~kpc \citep[e.g.][]{Jones1993,Oudmaijer1995}.

Here, we determine the distance to the source using the DR3 Gaia parallax measurements \citep{Gaia2021}.  Simply inverting the parallax to determine an object's distance is reliable as long as the error of the parallax over the parallax itself is small ($<$0.1). For larger relative errors, a simple inversion not only introduces more biases but is also known to perform poorly in treating non-linear effects. On the other hand, the method proposed by \citet{Bailer2021} adopts a Galactic model based on which a distance prior is used for the distance determination. In the case of IRC+10420 the measured parallax is 0.20$\pm$0.06~mas (implying a distance of 5.0$^{+2.1}_{-1.2}$~kpc), while the Bailer-Jones distance of d = 4.26$^{+0.88}_{-0.75}$~kpc\footnote{The calculation was retrieved from the Gaia archive, https://gea.esac.esa.int/archive/}, is a more accurate measurement and within the range of distances reported in previous studies.  We will continue with the distance of 4.26~kpc, which results in a luminosity of $\sim$4.6$\times$10$^{5}$~L$_\odot$ \citep[L $=$ 25462 (D/kpc)$^{2}$ ~L$_{\odot}$;][]{Oudmaijer2013}. Adopting the effective temperature of 7930~K \citep{Nieuwenhuijzen} results in a stellar radius of $\sim$360 R$_\odot$ (1.7~au at 4.26~kpc). We note that if we adopt an effective temperature of 8500~K \citep{Klochkova1997}, which is more in line with what is expected for A3/A6 type supergiants the resulted stellar radius is smaller by $\sim$13\% ($\sim$314 R$_\odot$; 1.5~au).

\subsection{Geometric modelling}
\label{modelling}

The GRAVITY data allow direct measurements of the sizes of the components we detect in the environment of the YHG. We identify three contributions (continuum, Br$\gamma$ and neutral gas) of which the geometry of the continuum and the circumstellar gas (Br$\gamma$ emission) were analyzed with much sparser data in OdW13. In this paper we are also able to constrain the geometry of the neutral gas via the doublet sodium emission for the first time. 

For the purpose of this analysis we adopt a polychromatic modelling approach, which takes into account the entire spectro-interferometric information. The tool we use is called PMOIRED and it is developed by Antoine M\'erand (see also Sect.~\ref{sec:obs}). The code uses parametric modelling to simulate the interferometric signals as produced from different brightness distributions, such as discs, gaussians and rings, and provides the flexibility for their combination as well. The best fit is then achieved by a least-square approximation and one can evaluate the fit by retrieving the bootstrapping estimation of the uncertainties. 

\subsubsection{K-band continuum}
\label{cont_geom}

The observed squared visibilities of the continuum obtained in the full spectral coverage of GRAVITY/VLTI (2-2.44 $\mu$m) show an overall decrease with increasing wavelength (Figure~\ref{fig:fit}, right panel), while the visibilities are always $<$ 1 even at shorter baselines. The low visibilities at short baselines indicate the presence of an extended emission component, while the wavelength dependence indicates that the shorter wavelength emission arises from a more compact region with respect to longer wavelengths. 

To constrain the size and geometry of the K-band compact emission, we perform fits using the $\chi^2$ minimization approach. We adopted several combinations of brightness distributions (gauss, disc, ring). The best fit (reduced $\chi^2$ $\sim$3) is achieved when adopting a gaussian distribution for the compact component, while adding an extra flux contribution of an extended/diffuse emission component with respect to the resolution and FoV of GRAVITY on ATs. We find that the best fit model (Figure~\ref{fig:fit}) results in a size of 0.86$\pm$0.08~mas (3.7~au at 4.26~kpc) for the compact component, while the flux contribution of the diffuse component is 22$\pm$2~\%. We note that the flux contribution of the diffuse continuum component is higher compared to what was reported in OdW13 (6\%). The observed increase can be result of the $\sim$3 times larger FoV achieved with GRAVITY on ATs compared to AMBER on UTs \footnote{We cannot rule out that part of the observed increase is a result of calibration systematics on absolute visibilities observed with AMBER using the FINITO fringe tracker (e.g., fringes jumps are known to be causing decreases in visibilities).}. 

The observed closure phases of the continuum are centered around 0$^{\circ}$, which means that the brightness distributions for both components is concentric and symmetric. We conclude that the 2.2 $\mu$m compact component of the continuum traces the photosphere of IRC+10420. In particular, the continuum emission has an observed radius of 1.8~au at the distance of the source, nicely consistent with our estimate of the stellar radius inferred from spectroscopy (R$_{*}$ $\sim$ 1.4-1.7~au, according to the values provided in Sect.~\ref{dist}). 

\subsubsection{Br$\gamma$ and neutral gas emission}

The geometry and size of the neutral gas component in the wind of IRC+10420 is determined for the first time. In this subsection, we compare both gas components as traced via the Br$\gamma$ and Na {\sc i} doublet emission. We follow a similar approach to that used for the continuum emission, adopting various combinations of brightness distributions (gauss, disc, ring). To limit the amount of free parameters during the fitting process of both lines, we fix the contribution of the continuum emission (size and geometry) to match the one obtained in Section~\ref{cont_geom} (FWHM $=$ 0.86~mas).  

\begin{figure*}
\begin{center}  
\includegraphics[scale=0.22]{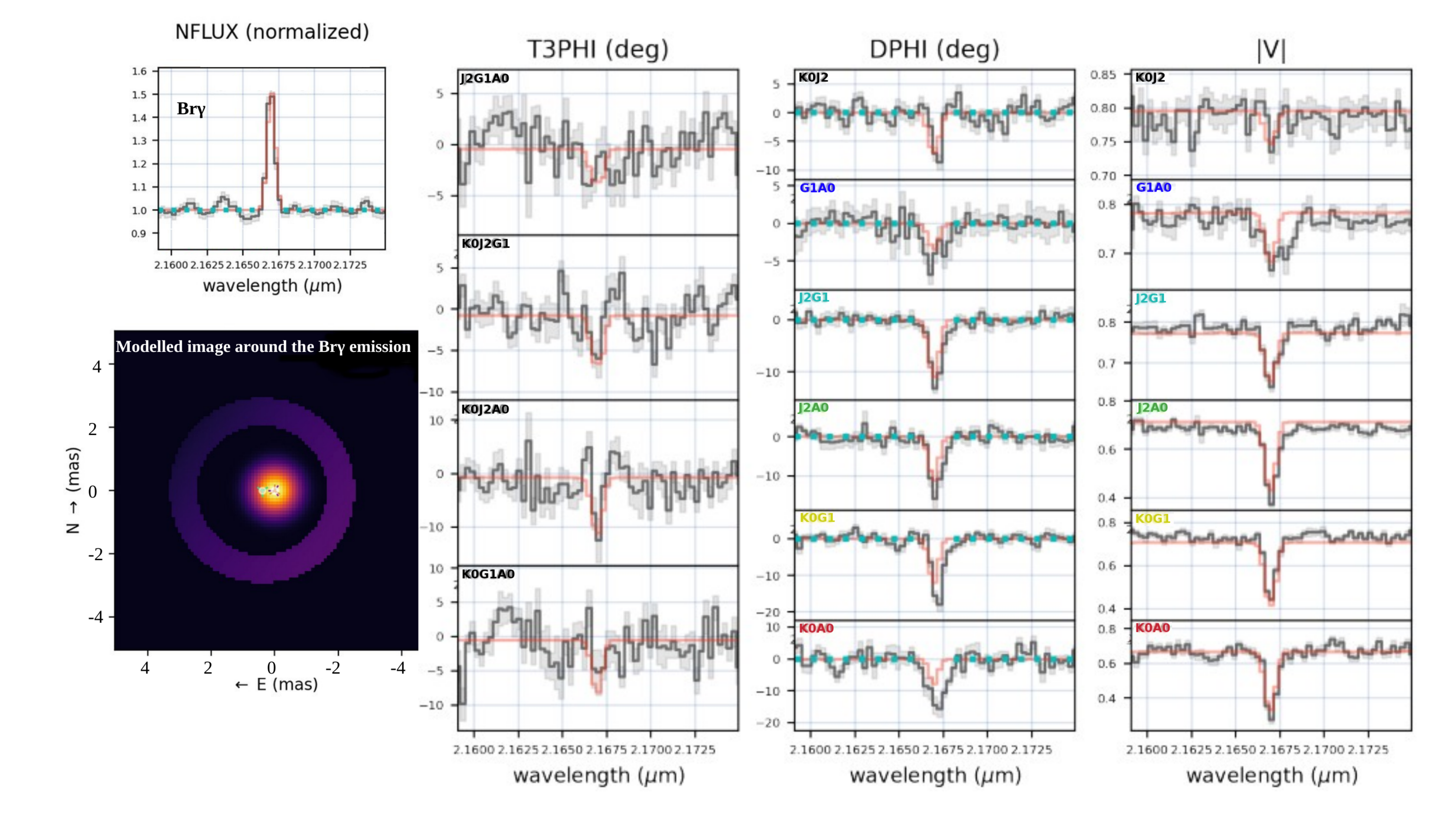} \\
\includegraphics[scale=0.22]{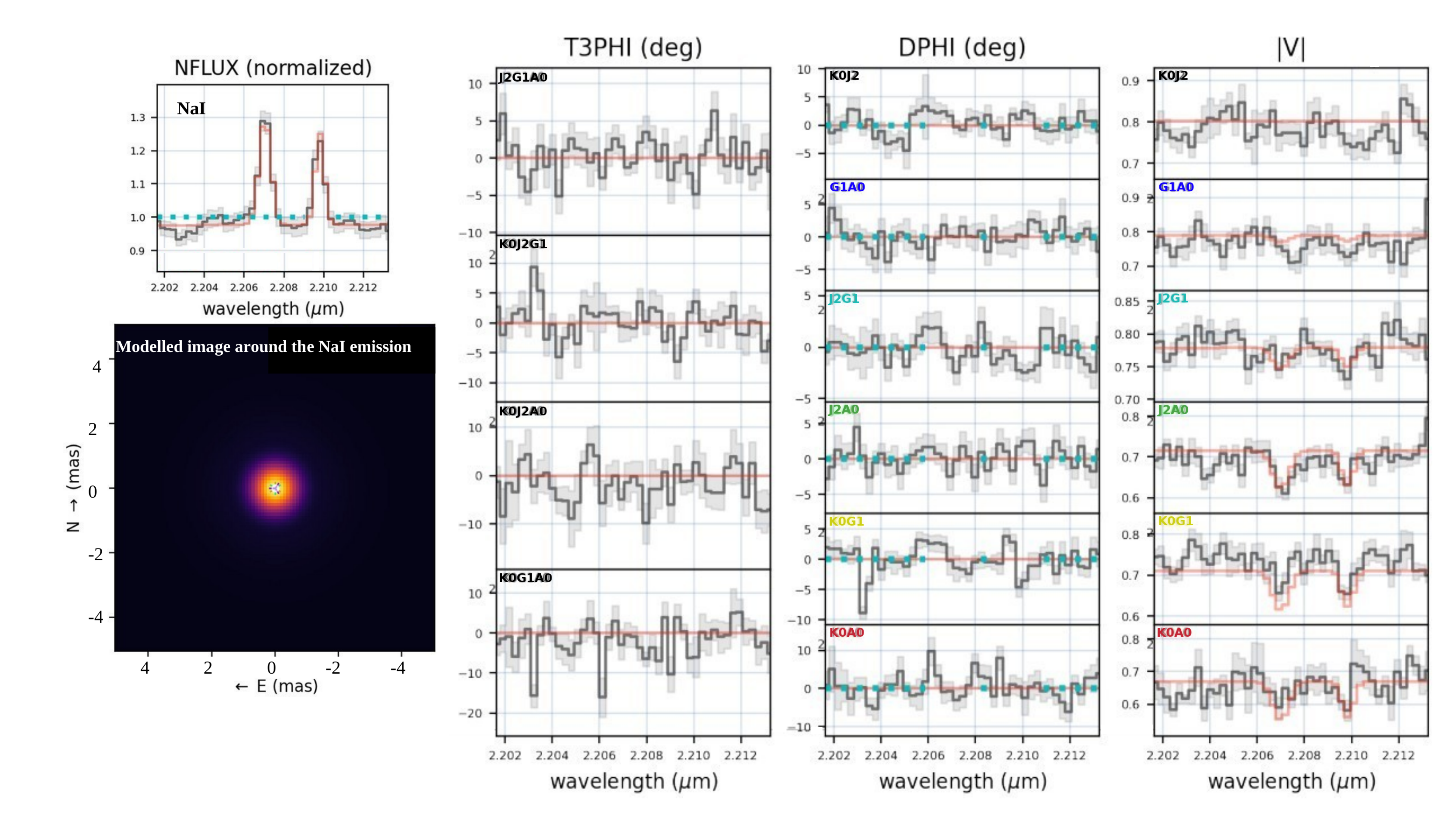}
\end{center}
\caption{Model image of IRC+10420 around the Br$\gamma$ and Na {\sc i} doublet emission (averaged) and continuum. The geometric models of the different emission components (red line) fit well all interferometric observables (closure phases; T3PHI, differential phases; DPHI, visibilities; V, and line profile; NFLUX; black line) on the targeted spectral coverage. The color scheme of the baselines follow the one presented in Figure~\ref{fig:uvplane}.}
\label{fig:model_image}
\end{figure*}

We find that the best fit of the Br$\gamma$ emission (reduced $\chi^2$ $\sim$ 2.8) can be achieved for a ring brightness distribution of a diameter, D $=$ 5.8$\pm$0.1~mas and width 0.3~mas. The Br$\gamma$ emission is characterised by significant changes in differential (-6$^{\circ}$.. -18$^{\circ}$) and closure phases (as low as -12$^{\circ}$) with respect to the continuum (Figure~\ref{fig:model_image}) which are indicative of diversion from symmetric emission and therefore require the introduction of a more complex brightness distribution. To account for asymmetries during the fitting process, we introduced a slant in flux along the ring at a projection angle of 53$\pm$6$^{\circ}$ (for the fainter part of the ring or 233$\pm$6$^{\circ}$ for the brighter part; 0$^{\circ}$ at North, 90$^{\circ}$ at East). In particular, the reduced $\chi^2$ dropped from $\sim$5 to 2.7-2.9 when a slant of 0.5-1 was introduced (for a value $=$ 1, the one side of the brightness distribution has a 0 flux). Lastly, we find that the closure phases are best fitted when there is a small offset, $\Delta$R.A., of 0.30$\pm$0.07~mas between the ring distribution and the central gaussian distribution of the continuum. When accounting for larger offsets the fit significantly worsens (reduced $\chi^2$ $>$ 10). We note that simpler geometrical models were tested but could not reproduce the interferometric observables as well. In particular, the best fit when we adopt a gaussian brightness distribution for the Br$\gamma$ emission, which is characterised by a shifted photo-centre compared to the continuum is found to be at an offset of $\sim$1~mas. In this case, the resulted reduced $\chi^2$ is systematically $>$ 3. Although the observed differential phases and visibilities can be reproduced at similar quality as when adopting the ring, the fit on closure phases worsens (from typical $\chi^2$ of 1.5-2 to typical $\chi^2$ of 1.8-3.8). 

To constrain the size and geometry of the Na {\sc i} doublet emission we follow a similar approach as that of the Br$\gamma$ emission. We find that the best fit (reduced $\chi^2$ $\sim$ 2.2) is achieved for a simple gaussian brightness distribution of FWHM $=$ 2.9$\pm$0.1 mas. We note that adopting a disc or a ring worsens the fit to the interferometric observables (reduced $\chi^2$ $>$ 4.3). Figure~\ref{fig:model_image} demonstrates that both the differential and closure phases around the Na {\sc i} doublet emission do not show significant variations with respect to the continuum ($\sim$0$^{\circ}$), which is indicative of the symmetric nature of the sodium emission with respect to the continuum. The symmetric nature of Na {\sc i} is also supported by the fact that the Br$\gamma$ emission does show variations in phases even though it is larger and more resolved than Na {\sc i}. Therefore, there is no necessity in introducing more complex brightness distributions or asymmetries during the fitting process. 

Our geometrical models demonstrate that both the Br$\gamma$ and the Na {\sc i} doublet are more extended compared to the continuum emission. In addition, Na {\sc i}  emission originates from a layer closer to the central star and it is more symmetric compared to the Br$\gamma$, with the latter showing a non-homogeneous ring brightness distribution similar to that reported in OdW13.

\subsubsection{Spatial measurements of the hot emission within a decade}

The measurement of the continuum size using GRAVITY data (2.2$\mu$m continuum emission) is not the first measurement of the star's size by means of NIR interferometry. In particular, we find it to be within the estimated size range reported in OdW13 based on AMBER observations taken in 2009 (0.7~mas-0.98~mas). OdW13 provided a distance independent expectation of the stellar diameter of IRC+10420 to be $\sim$0.7~mas, therefore our modelling suggests that when accounting for errors the K-band continuum is up to 10\% larger than the size of the photospheric emission of IRC+10420. Moreover, our findings indicate that the compact continuum emission within those 7 years originates from a stable environment around IRC+10420 without signs of an expansion or an infall. In particular, at the angular resolution of the observations any steady contraction or expansion motions with velocities of order of only a few kms$^{-1}$ ($\sim$ $>$ 1.5~kms$^{-1}$) should be spatially traceable in the observed time-frame of 7 years. We note that weak shock waves associated with yellow hypergiants are of order of 0.5~kms$^{-1}$ \citep{Lobel2001} and therefore would remain unseen, but pulsations \citep[1-4 kms$^{-1}$;][]{vanGenderen2019} or eruptive motions \citep[30-100~kms$^{-1}$; e.g.][]{Koumpia2020} should be detectable. In summary, the size of the continuum emission appears to have remained constant during the 7 years between the NIR interferometric observations.

We now investigate the size and geometry of the Br$\gamma$ emission during this period.  Similarly to previous studies, Br$\gamma$ line emitting region is found to deviate from spherical symmetry \citep[see also, OdW13,][]{Driebe2009}. In OdW13 it is suggested that ring brightness distribution is a projection of an hour-glass wind geometry, seen close to pole-on \citep[see also HST;][]{Tiffany2010}. 

In particular, OdW13 argued that the system needs to be inclined by 8$^{\circ}$ in order for the star to be in a position to obscure parts of the wind as traced by the Br$\gamma$ emission. The determined ring diameter is $\sim$ 40\% larger than that presented in OdW13 (4.18~mas). Our latest observations reveal that in a 7 year interval the ring has increased by 0.8~mas on the sky, which at the distance of the source (4.26~kpc), corresponds to a velocity of 2.5 km$^{-1}$. This estimation is if we assume a shell-like structure surrounding the central star. Based on the specific geometry of the Br$\gamma$ emission though (i.e., hour glass), a more meaningful measurement is that of the line-of-sight outflow velocity, which we estimate as follows. 

\begin{figure*}
\begin{center}  
\includegraphics[scale=1]{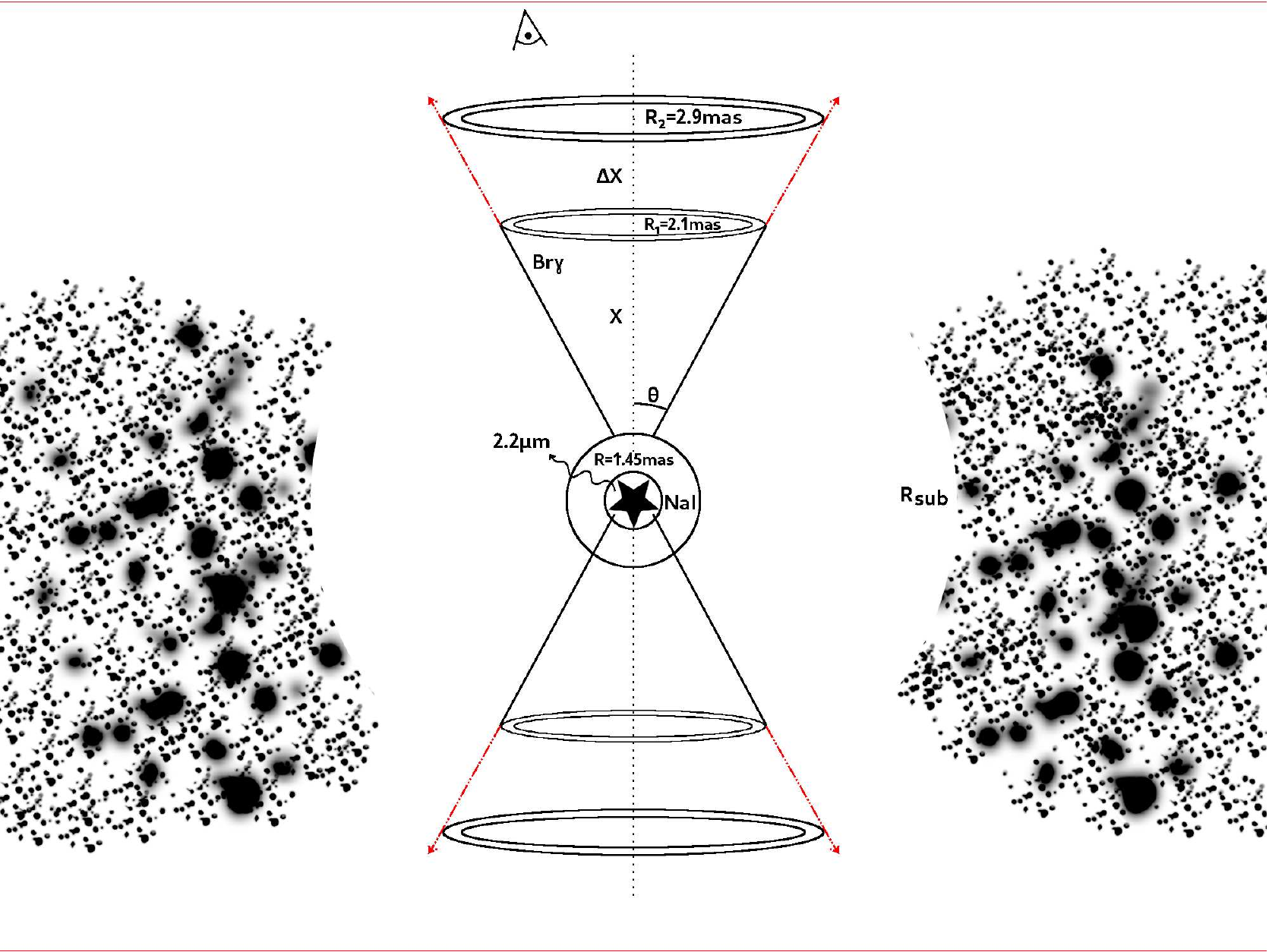}
\end{center}
\caption{Schematic view of the modelled geometry of the gas emission around IRC+10420 as traced with Br$\gamma$ and Na {\sc i}. The 2.2 $\mu$m compact emission originates from the stellar photosphere, while Na {\sc i} is a bit more extended. We observe an increase in the ring associated with Br$\gamma$ in a timescale of 7 years, which is also demonstrated in the sketch. For reference we also show the dust sublimation radius ($\sim$~5mas). The observer is represented by the eye symbol on the top and looks at the geometry almost face-on (8 $^{\circ}$). The sketch is a scaled version of the actual geometry with the exception of the opening angle which for illustrative purpose is wider than the one we determine in this study (3-10$^{\circ}$)}.
\label{fig:sketch}
\end{figure*}

In Figure~\ref{fig:sketch} we graphically illustrate a scaled version of the geometry of the close environment of IRC+10420, including the dust sublimation radius, the neutral gas zone (Na {\sc i}), and the hour-glass geometry of the Br$\gamma$ emitting region which indicates an evolution in the ejecta in a timescale of 7 years. If we look at the Br$\gamma$ emission at the edge of the cone, we find that the Br$\gamma$ emitting zone has "travelled" along the line-of-sight (or a propagated shock has reached hydrogen gas at) an extra distance of $\Delta$X = X/2.57\footnote{For an expansion towards the line of sight: $\Delta$X = X $\times$ [(R$_{2}$/R$_{1}$) - 1] $\times$ cos{\it{i}}, where R$_{2}$ = 2.9~mas, R$_{1}$ = 2.1~mas and {\it{i}} = 8 $^{\circ}$ (inclination with respect to the line of sight).} (where tan$\theta$ = R$_{1}$/X) in 7 years. We note that the opening angle has a high impact on this calculation. For this class of objects the outflow velocities are expected to be between 30~kms$^{-1}$ - 100~kms$^{-1}$ \citep[40~kms$^{-1}$ for IRC+10420 based on CO spectral observations;][]{Oudmaijer1996}, which can be reproduced for opening angles 2$\theta$ between 3 and 10$^{\circ}$.

Therefore, in this study we set an important constrain on the opening angle of the hour-glass geometry associated with the Br$\gamma$ emission, which is significantly smaller than what was previously assumed (45$^{\circ}$; OdW13). The small opening angle can explain the observed asymmetry in the brightness distribution of the ring, as a part of the receding side of the hour-glass shaped wind could be blocked by the star, although the presence of clumps can be an alternative explanation.   

Lastly, we examine possible variations of the Na {\sc i} emission. Our spectro-interferometric models show that Na {\sc i} originates from a region within that of the dust sublimation radius (i.e., no dust shielding mechanism). Also we find that this neutral emission is half the size of the Br$\gamma$ emission and therefore originates from a region closer to the star. A projection effect, where the sodium emission originates from a "blob" along the line of sight, cannot be excluded but it would require a very specific/ideal geometry which makes this scenario not very probable. We note that the dust sublimation radius of IRC+10420 is expected at 22~au (i.e. $\sim$5~mas at 4.26 kpc) if one assumes a dust sublimation temperature of 1500~K, therefore both the Br$\gamma$ and the Na {\sc i} originate from a region free of dust but which is more extended than the stellar photosphere.


\section{Discussion}

IRC +10420 is one of the very few YHGs known to be in the post-RSG phase defining the upper luminosity boundary in the HRD. This source is well studied with in different wavelengths and techniques which combined show a very complex mass-loss history of the source, with multiple shell ejections which are characterised by jet-like structures, rays, knots and arcs, and enhanced mass-loss within the past 400 years \citep[e.g., HST;][]{Humphreys1997}. In addition it provides an excellent opportunity to study the hotter regions close to the central object since it is verified by multiple studies to be close to pole-on.  Below we discuss our findings on the spectral and spatial morphology and variability of IRC+10420 in a timescale of decades.

\subsection{Spectral variability}

We presented the first blue optical spectra of IRC +10420, since 1994. The data enabled us to determine that the spectral type  did not, or hardly, change over 25 years.  IRC +10420 has not always been so stable regarding its spectral type. In particular, only a couple of decades ago, based on spectral variations it was argued that in a time-scale of $\sim$25 years its stellar spectral type has changed from late F-type to mid-A, which suggests an effective temperature jump of $\sim$2000~K (from 6000~K to 8000~K) in the corresponding timescale \citep{Oudmaijer1996,Oudmaijer1998,Klochkova2002}. This is in accordance with more studies suggesting that IRC+10420 is a post-red supergiant which follows its blue-ward path of evolution on the HR-diagram \citep{Jones1993}. Here, we report that the spectral type (and thus its temperature) has not, or hardly changed since then. Therefore, it would appear that this yellow hypergiant has "settled" for the time being, and has possibly "hit" the "White Wall" in the HR-diagram, preventing it from evolving further to the blue part \citep{dejager1998,Nieuwenhuijzen}. This is a critical phase of massive stellar evolution and a heavy burst of mass-loss in the near future is to be expected \citep{Stothers2001}.




\subsection{Spatial morphology}

The geometric modelling confirms an axi-symmetric wind origin of the Br$\gamma$ emission in a shape of a non-homogenous ring brightness distribution along the P.A. of $\sim$ 53$^{\circ}$. OdW13 first introduced a similar ring geometry based on AMBER observations and attributed it to the projection of an hour-glass as seen at an inclination of 8$^{\circ}$.
The (near) pole-on orientation had been found by a variety of techniques in previous works and has become well established \citep{Tiffany2010,Shenoy2015}.  
The hourglass geometry has been confirmed by
interferometry which traced the Br$\gamma$ emission around the star (OdW13) and by high-resolution adaptive optics assisted 2.2 $\mu$m imaging polarimetry \citep{Shenoy2015}.

Here we compare the determined wind geometry seen at milli-arcsecond to the findings of previous studies up to arcsecond scales and discuss the possible causes of an hour-glass shape.  During the past decades evidence for the presence of an axi-symmetric wind associated with IRC+10420 at various scales has accumulated. On arc-second scales the HST optical images revealed an axi-symmetry in the reflection nebula along the North-East/South-West direction \citep[P.A. $\sim$ 33 $^{\circ}$;][]{Humphreys1997}, while the infrared images of similar scales point towards an axi-symmetric emission along 58$^{\circ}$. At smaller scales of only a few stellar radii, spectropolarimetry probes an asymmetry along the short axis of the extended reflection \citep[almost perpendicular to what has been reported so far;][]{Patel2008}. \citet{Davies2007} also argue that wind axi-symmetry is real, based on the equivalent width and velocity variations of emission lines (H$\alpha$, Fe {\sc ii}) along the long axis of the reflection nebula (Southwest-Northeast).  

A possible mechanism responsible for the observed axi-symmetry of the wind is that of a rotation. Rotation is known to play a crucial role in the evolution and therefore morphology of massive stars which can also explain axi-symmetric wind geometries due to the non-constant effective gravity between poles and equator \citep{Maeder2000}. The rotation can be due to an intrinsic property of the star or due to the presence of one or more companions. In fact bipolar and hourglass shape geometries are seen in the gas surrounding LBVs \citep[$\eta$ Car, AG Car][]{Weis2013} or post-AGB stars \citep[e.g. HD 101584;][]{Olofsson2019}, where the presence of a companion is known to drive the characteristic shape. Our high angular resolution K-band interferometric observations with GRAVITY do not show characteristic binary signatures in either the visibility or closure phases (nor do the spectra), therefore, we can possibly exclude the presence of a companion at separations $\sim$7-800~au at the traced position angles. We note that given the incomplete uv-coverage of the observations, we cannot trace about half of the possible binary configurations between $\sim$7-45~au. Moreover, on the search of a companion, PMOIRED finds a 3$\sigma$ detection limit of 3.7 mag and a 2$\sigma$ detection limit between 2.8 and 4.1 mag. In addition, the spectra would not be able to reveal a companion orbiting the primary star in an orbital plane close to pole-on. We conclude that our dataset cannot exclude the presence of a companion located at au separations, while the possible different evolutionary stage of the system also makes an observational confirmation challenging \citep[i.e., different expected distances and mass for the companion, see HD~101584;][]{Kluska2020}.  

\section{Concluding Remarks}

This paper presents new NIR high angular resolution interferometric observations (GRAVITY/VLTI) tracing material down to mas of the YHG IRC+10420. We investigate spectral and spatial variability and report on the geometry of the 2.2~$\mu$m continuum emission and that of the Br$\gamma$ and the neutral Na~I line emission. 

\begin{itemize}

\item The 2.2~$\mu$m continuum emission has maintained its size during the 7 years spanning the VLTI observations.  Its size is consistent with a photospheric origin.
\item Br$\gamma$ shows an asymmetric ring brightness distribution with respect to the continuum at a position angle of $\sim$53$^{\circ}$ traced by variations in both differential and closure phases. Our independent findings support the scenario of an hour-glass wind geometry as previously proposed. 
\item The wind (traced by Br$\gamma$) is characterised by a ring shape which has expanded by 40\% over 7 years. Accounting for an outflow velocity between 30~kms$^{-1}$ and 100~kms$^{-1}$ we are able to set a constraint on the opening angle (2$\theta$) of the hour-glass geometry to be between 3$^{\circ}$ and 10$^{\circ}$.
\item {Our interferometric K-band GRAVITY observations do not reveal the presence of a companion at 7-800~au separations (at least at the achieved contrast limit of 3.7~mag at 3$\sigma$) that could potentially explain an hour-glass morphology of the wind.}
\item Both the Br$\gamma$ and the Na {\sc i} emission originate from a region inside the dust sublimation radius and are larger than the continuum (photospheric) emission.
In addition, the Na {\sc i} emitting region is symmetric and more compact compared to the Br$\gamma$ emission. 
\item We report no significant spectral variability over the last 25 years, suggesting a stable phase for IRC+10420, in which its temperature has essentially remained  unchanged. This is in contrast to what had been observed previously when the observed temperature was found to have increased by 2000K in temperature over 20 years.  This suggests that this Yellow HyperGiant has "hit" the white wall in the HR-diagram preventing it from evolving even further blue-wards.

\end{itemize}

 This paper provides spectroscopic evidence that IRC+10420 experiences a halt in its temperature evolution (within a 25 year timescale), which can also explain the "plateauing" observed in J-band photometry of the source \citep{Patel2008}. IRC+10420 is therefore a very unique object, which gives us the opportunity to witness a yellow hypergiant hitting the white wall / yellow void or reaching its maximum red-ward shift as predicted by the bistability mechanism. At present, any further spectral evolution is halted, therefore studies can take advantage of this unique phase in massive stellar evolution and focus on tracing variations of mass-loss geometry and rates traced by excited/ionised gas at small au scales (e.g. through H$\alpha$ and Br$\gamma$ spatial and spectral variations). In particular, the dynamical instability that a YHG will experience once hitting the white wall / yellow void while moving blue-ward is predicted to come along with very high mass-loss rates \citep{dejager1998}, and this is something future interferometric observations (K-band, M-band) and modelling will be able to confirm. 
 
 Our geometrical models show that the Na {\sc i} emission stems from a gas layer which is located closer to the central star compared to the Br$\gamma$ emission. If Br$\gamma$ emission is a result of recombination, and therefore, traces an ionised medium, the observed relative sizes of the Br$\gamma$ and Na {\sc i} emission contribute to a rather puzzling observation. This was previously addressed towards another object of this rare class, IRAS 17163$-$3907 \citep{Koumpia2020}. Future modeling can address this long-standing enigma by investigating if a two-zone model in strict local thermal equilibrium, rather than the popular scenario of a pseudo-photosphere, can explain both the K-band emission spectrum and the sizes of the various line emission components of IRC+10420 and other known objects of this class. In addition, our geometrical models independently confirm that the wind is better represented by a ring-like brightness distribution pointing towards the hour-glass geometry proposed by OdW13. Future interferometric and radial velocity observations at shorter wavelengths (e.g. PIONIER) are needed to better investigate if the presence of a companion at the inner au scales is responsible for the observed axi-symmetric shape. Finally, it is of great importance to monitor IRC+10420 both photometrically and spectroscopically, since all the signs indicate that this yellow hypergiant will evolve to become a Wolf-Rayet or undergo a large supernova outburst.

\section*{Acknowledgements}

E.K. was funded by the STFC (ST/P00041X/1).
Based on observations made with the William Herschel Telescope
operated on the island of La Palma by the Isaac Newton Group of
Telescopes in the Spanish Observatorio del Roque de los Muchachos of
the Instituto de Astrof\'{i}sica de Canarias. The ISIS spectroscopy
was obtained as part of SW2018a27. Based on observations collected at the European Southern Observatory under ESO programme(s) 60.A-9434(A) (X-Shooter), 077.D-0729(A) and 091.D-0221(A) (FEROS) and 60.A-9166(A) (GRAVITY). This project has received funding from the European Union's Framework Programme for Research and Innovation Horizon 2020 (2014-2020) under the Marie Sk\l{}odowska-Curie Grant Agreement No. 823734. 

\section*{Data Availability} 
The reduced data presented in this paper will be shared on reasonable request to the corresponding author.




\bibliographystyle{mnras}
\bibliography{example} 




\appendix

\section{GRAVITY observations of IRC+10420}

\begin{figure*}
  \includegraphics[scale=0.7]{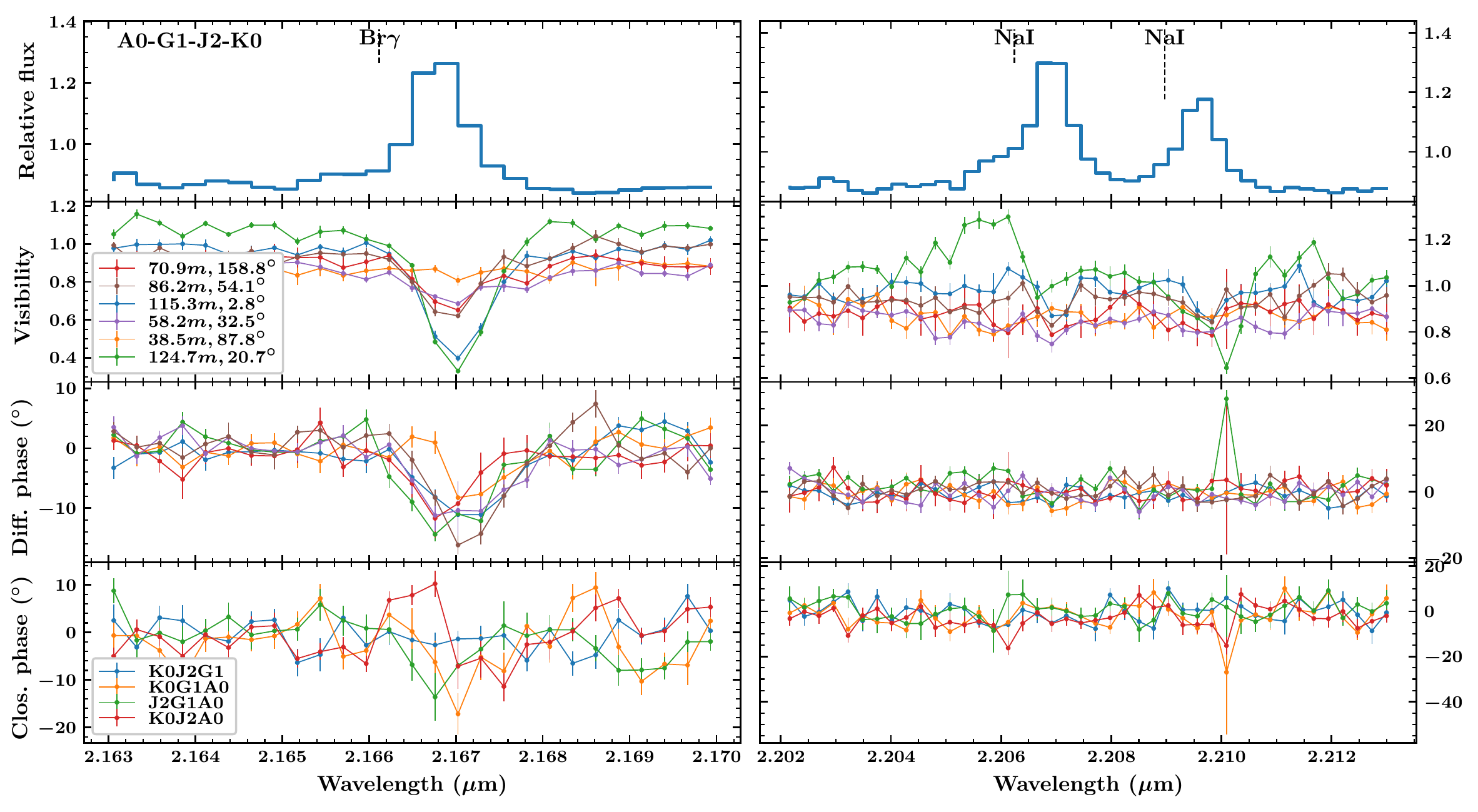}
\caption{Large configuration: Relative flux, visibility, differential and closure phase as a function of wavelength around the Br$\gamma$ and Na {\sc i} doublet emission towards IRC+10420 (1 observing run). The 6 baselines and 4 triplets of A0-G1-J2-K0 configuration are plotted with different colors. Both Br$\gamma$ emission and Na {\sc i} show a smaller visibility than the continuum while Br$\gamma$ shows a larger drop. A differential phase change is also detected towards Br$\gamma$ but not towards Na {\sc i}. The closure phases of Na {\sc i} do not show changes.}
\label{fig:vis_brg1_2_irc}
\end{figure*}

\begin{figure*}
\begin{center}  
\includegraphics[scale=0.4]{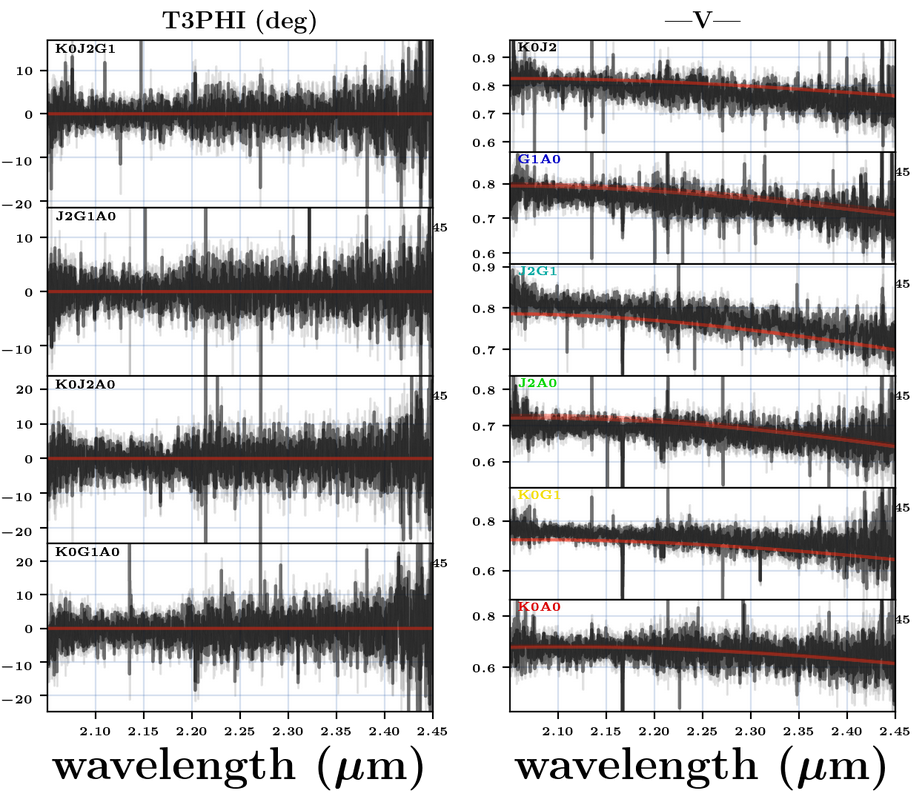} 
\end{center}
\caption{Observed visibilities and closure phases of the continuum towards IRC+10420 as measured with GRAVITY, plotted as a function of wavelength for each baseline of the large configuration (black). The predictions of the best fit model using PMOIRED are also overplotted in red.}
\label{fig:fit}
\end{figure*}









\bsp	
\label{lastpage}
\end{document}